# Finding and Characterizing Other Worlds: the Thermal-IR ELT Opportunity

Michael R. Meyer (University of Michigan), Thayne Currie (NOAJ), Olivier Guyon (NOAJ), Yasuhiro Hasegawa (JPL), Markus Kasper (ESO), Christian Marois (NRC-CNRC), John Monnier (U. Michigan), Katie Morzinski (U. Arizona), Chris Packham (UT-San Antonio), and Sascha Quanz (ETH)

*Executive Summary:* The next generation ground-based extremely large telescopes (ELTs) present incredible opportunities to discover and characterize diverse planetary systems, even potentially habitable worlds. Adaptive-optics assisted thermal-IR (3-14 $\mu$m) imaging is a powerful tool to study exoplanets with extant 6-12 meter telescopes. ELTs have the spatial resolution and sensitivity that offer an unparalleled expansion of the available discovery space. AO-assisted thermal-IR instruments on ELTs will be superior to JWST for high contrast imaging in the thermal-IR, and complementary to high contrast observations at shorter wavelengths, in space or with second-generation extreme AO instruments. With appropriate investments in instrumentation and pre-cursor observations, thermal-IR equipped ELTs could image the first terrestrial and super-earth planets around nearby stars, opening the door to characterization of potentially habitable planets from the ground and space.

Direct imaging and secondary eclipse measurements are two ways to directly detect light from other worlds. To date, direct detection is limited to dozens of objects beyond several AU (high contrast imaging – see Figure 1) and within a few tenths of AU (secondary eclipse). While characterizing these atmospheres provides fundamental insights, most exoplanets are found at intermediate separations which have yet to be studied (e.g. Meyer et al. 2018, A&A, in press, arXiv:1707.05256). Characterizing those within the liquid-water or 'habitable zone' of stars is clearly a priority. The probability for transit, and thus secondary eclipse, drops as a function of semi-major axis, while direct imaging of self-luminous planets is easier at larger orbital radii.

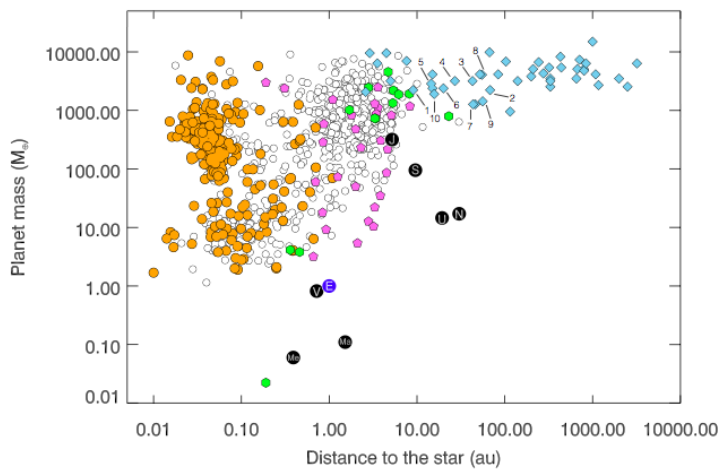

*Figure 1 - Planet mass as a function of orbital separation (from Pepe et al. 2014, Nature, 513, 358). Orange circles are transiting planets (including those with direct detection secondary eclipse) with confirmed RV masses. Open circles are RV discovered planets, while green points represent planets detected through timing variations. Magenta points represent microlensing detections and blue diamonds are companions discovered through direct imaging (those numbered have mass ratios relative to their star < 0.02 within 100 AU). Notice the gap between transits (including secondary eclipse) in orange and direct imaging in blue.*

Imaging can be achieved by observing reflected light or thermal emission each carrying important characterization information. Both reflect light and thermal emission of a planet depend on stellar luminosity, orbital separation, planet radius, and albedo as a function of wavelength: the amount of reflected light is high when the albedo is high while the thermal emission is highest when the albedo is low. In the context of obtaining estimates of effective temperature and luminosity, thermal emission is crucial. In addition, direct detection may offer the only estimates of planet radius for non-transiting planets. Planets whose energy budget are dominated by their heat of formation are hotter and brighter when young, providing powerful constraints on theories of formation and early evolution. Planets in thermal equilibrium with their host star emit steady amounts of infrared radiation regardless of age. Wavelengths < 2.3 μm are best suited for imaging planets in reflected light as well as thermal emission for the youngest (< 30 Myr) and hottest (> 800 K) planetary mass companions. Thermal infrared wavelengths from 3-14 μm are best-suited for imaging planets with temperatures from 270-800 K, including those in the liquid-water zone.

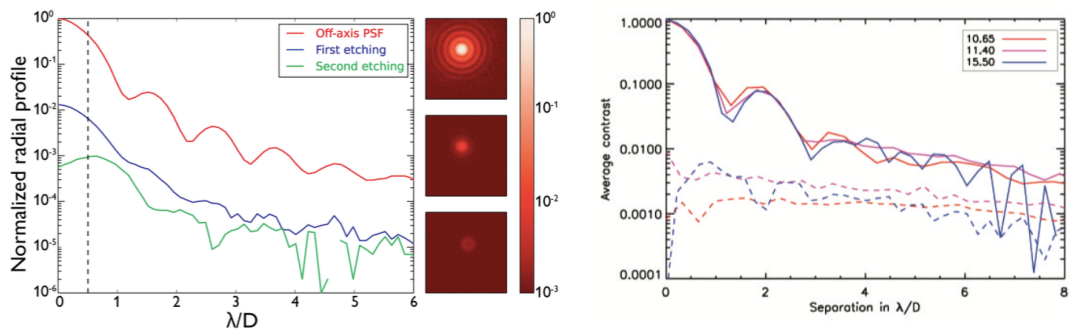

*Figure 2 - At left raw contrast achievable with a thermal-IR annular grove phase mask coronagraphs in the lab (Delacroix et al. 2013, A&A, 513, A98). At right the achievable contrast with the MIRI four quadrant phase mask on JWST (Boccaletti et al. 2015, PASP, 127, 633). AO-assisted thermal-IR imaging from the ground should provide superior contrast out to at least 8 λ/D.*

What is the role of thermal-IR ground-based telescopes in the era of JWST? The challenge of exoplanet imaging is not just one of sensitivity at the desired wavelengths, but also spatial resolution and achievable contrast. Indeed, it is expected that even existing 6-12 meter ground-based telescopes (equipped with adaptive optics and diffraction suppression optics) will out-perform JWST in the contrast limit (see Figure 2). Shorter wavelengths are naturally limited by diffraction at the finest spatial scales ($\Theta \sim \lambda/D$, e.g. 0.065" at 1.65 μm on a 6.5 meter telescope). Yet achieving this ultimate resolution is easier at longer wavelengths where wave-front errors are more manageable. We have yet to achieve a completely fair head-to-head comparison in achievable contrast for optimized extreme AO systems, but to date, the contrast achieved at fixed $\lambda/D$ is comparable. This implies finer angles resolved at shorter wavelengths. However, in imaging giant planets, this advantage is overcome by nature: planets that emit copiously at shorter wavelengths are warmer. Because such planets cool as they age, the required targets are young, and young stars are rare. Thus, samples for study are only found at greater distances, yielding comparable physical resolution from the finer angular resolution. There is always an age, above which, planets are colder, and easier to detect at longer wavelengths in thermal emission, even with the larger background from the sky and telescope. Older planets are found around older stars, which are more common, and therefore closer yielding physical resolution in the thermal-IR comparable or better to shorter-wave surveys.

Neither approach is fundamentally better, but complementary: depending on the distance to appropriate star samples, of a given age, the aperture of the telescope and quality and nature of the instruments available dictate which waveband is best for a particular survey. Indeed, the diagnostic power of spectra in the near-IR has been vital to further understanding these young gas giants (e.g. Bonnefoy et al. 2016, A&A, 587, 58). Nonetheless, it is a fact that most directly imaged young planets at large orbital radii are easier to detect in the L-band, though detailed characterization benefits tremendously from observations over the broadest wavelength range possible. For example, researchers have found evidence for clouds and/or non-equilibrium chemistry by comparing models to data from 1-5 $\mu$m (e.g. Skemer et al. 2014, ApJ, 792, 17). There are many important molecules that carry volatile species perhaps vital to the emergence of pre-biotic compounds ($CH_4$, $H_2O$, CO, HCN, and $NH_3$) in the thermal-IR. In particular, $NH_3$ is crucial to constraining the nitrogen abundance and traces "cool" chemistry with strong features in the 10 $\mu$m region. Broad wavelength coverage will permit detailed studies of cloud properties (particle size and composition) as their optical properties transition from gray to wavelength dependent in the thermal-IR. While JWST will have tremendous sensitivity in the background limit of space, it lacks the spectral resolution needed for important observations of molecules – the astrophysics of the cold and slow - where natural linewidths and relative velocities require R > 50,000 for optimum detection. The "thermal-IR advantage" increases with telescope aperture as one is able to detect colder planets (Heinze et al. 2010, ApJ, 714, 1570). It can even be realized at modest altitude sites. During times of low atmospheric water vapor (e.g. best 10-20 % at sites below 8000 feet), one is expected to be telescope emissivity limited in specific bands where the atmospheric transmission is > 90 % (e.g. 3.6-4.0, 8.6-9.3, and 9.9-13 $\mu$m), excellent bands to discover and characterize exoplanets from the ground. There are several 3-5 $\mu$m AO-assisted camera on 6-12 meter telescopes and currently none operating between 8 and 14 $\mu$m. The ESO-NEAR system will begin to fill this niche in 2019 and two other systems are under development for Magellan and Gemini.

This sets the stage for the ground-based ELT opportunity in direct imaging. With telescopes of diameter 24.5 to 39 meters, astronomers will have the sensitivity, and the angular resolution, to image orders of magnitude more planets compared to existing facilities over a wide range of sizes and orbital separation. In the thermal-IR, METIS will be one of the first three instruments on the E-ELT, and MICHI/MIISE are proposed second generation instruments for TMT/GMT respectively. Near-IR instruments utilizing extreme AO are planned in the second generation of instruments for all three ELTs. Of central importance will be to verify models of planet luminosity and temperature as a function of planet mass and age. ELTs will image dozens of planets with dynamical masses from RV as well as Gaia astrometry, providing the ground-truth needed to interpret demographic studies from imaging surveys (see Figure 3).

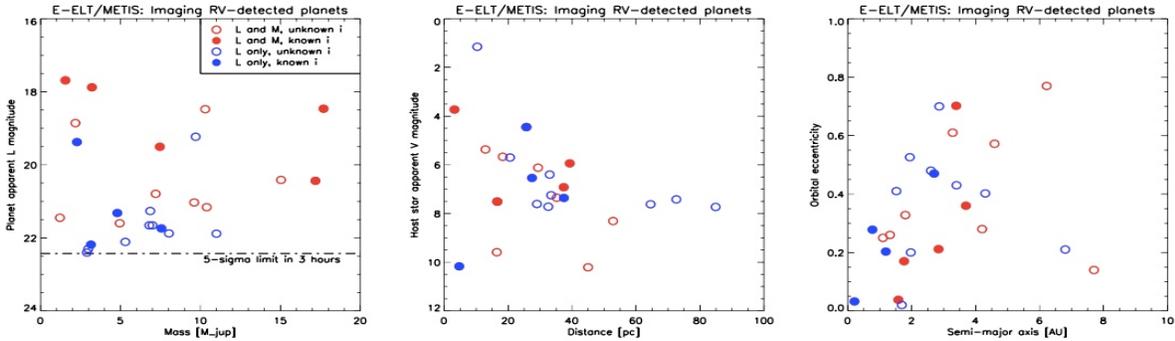

*Figure 3- Gas giant planets from published RV surveys thought to be accessible for direct imaging photometry and spectroscopy with the METIS thermal-IR camera on the 39 meter E-ELT (from Quanz et al. 2015, IJAsB, 14, 279). Additional candidates will be available from legacy RV surveys as well as astrometric detections with Gaia.*

Around nearby stars of intermediate age (< 50 pc and < 1 Gyr old) ELTs can conduct surveys, probing masses down to 0.1 Mjupiter (near the break in power-law mass function from gas giants to ice giants), and cross the peak in the surface density distribution of planets between 3-20 AU around sun-like as well as lower mass stars. Combined with demographics from RV, transit, microlensing, and astrometry, such studies will enable us to dissect exoplanet populations as a function of host star mass, confronting theories of planet formation and the dynamical evolution of planetary systems. Finally, ELTs will be able to image planets down to super-earth and terrestrial sizes around the very nearest stars both in reflected light and thermal emission in equilibrium with the isolation of the host star (cf. Hinz et al. 2009, arXiv:0902.3852). Extrapolating the results from Kepler as a function of planet size and orbital separation, combined with predictions of what a thermal-IR camera on an ELT can do, one can predict the likelihood that there is a rocky planet to be seen around the nearest sun-like and intermediate mass stars within 10 parsecs. The expectation value of this thought-experiment suggests that several will be imaged. Most of these will be warm super-earths but it is even possible to image a planet in the habitable zone of the very nearest stars (see Figure 4). Characterizing the atmospheres of these objects, even warm ones inside the liquid-water zone, will enable comparison to close-in super-earths studied through transit and secondary eclipse exploring possible evolutionary differences.

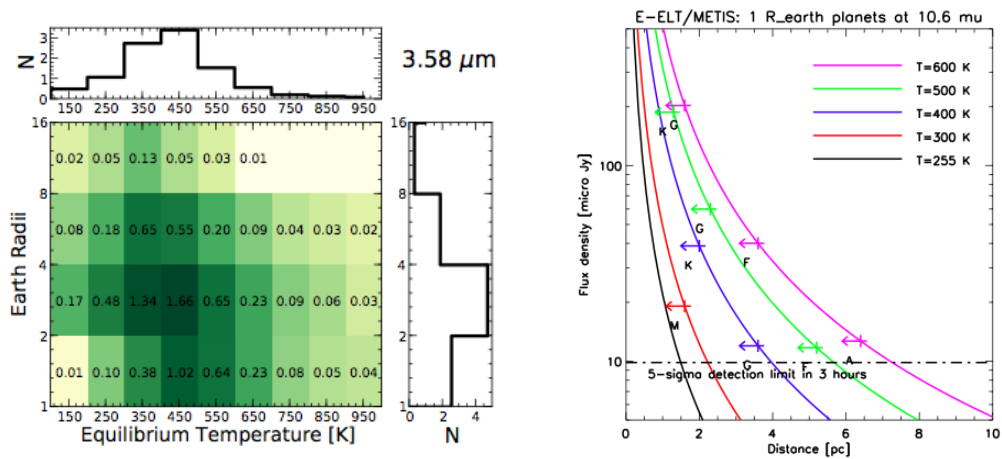

*Figure 4 - Detectability of terrestrial planets and super-earths from the expected performance of METIS on the E-ELT (from Quanz et al. 2015, IJAsB, 14, 279). At left are the expectation values of planets detected as a function of equilibrium temperature and planet radius (with marginal distributions for 3.58 mm L-band). At right are the estimated flux densities for one earth-radius planets as a function of planet temperature and distance in parsecs assuming thermal equilibrium.*

There are many powerful synergies between high contrast imaging in the visible/near-IR and AO-assisted thermal-IR imaging from 3-14 µm with ELTs. Perhaps the greatest would be to detect a terrestrial planet or super-earth in both reflected light and thermal emission, first through photometry and later spectroscopically. This would enable resolving the radius-albedo ambiguity, comparing energy received by the planet to energy emitted, and perhaps detection of an atmospheric greenhouse. Indeed, there is overlap between targets accessible to ELTs with thermal-IR imaging and those observable with WFIRST-AFTA. Similar samples will also be accessible with second-generation near-IR instruments with extreme AO systems on ELTs. Characterizing diverse rocky planet atmospheres with ELTs could be a capstone achievement for these extraordinary facilities, perhaps even capable of detecting biosignatures in the thermal-IR (e.g. Domagal-Goldman et al. arXiv:1801.06714). This later possibility could be enhanced through high dispersion spectroscopy coupled with high spatial resolution enabling the highest achievable contrasts (cf. Lovis et al. 2017, A&A, 599, 16). Finally, we note that knowing where to look would greatly facilitate the abilities of ELTs to directly detect terrestrial and super-earths around nearby stars. Precision RV (PRV) legacy surveys, of several years duration, could be a powerful set of precursor observations insuring success in the search to image rocky planets around neighboring star systems.

Thermal-IR instrumentation on ELTs presents tremendous opportunities for exoplanet discovery in the next 10-15 years. We have to plan now in order to realize these opportunities: 1) We must fund ELTs to which a wide range of astronomers can have access and those facilities should have thermal-IR capabilities and adaptive optics that enable fundamental exoplanet science; 2) Sustained funding for development in thermal-IR detectors, IR focal plane wave-front sensing, diffraction control and suppression, and novel processing algorithms is crucial; 3) Complementary extreme AO near-IR imaging on ELTs as well as near-term space-based facilities (such as WFIRST-AFTA) are also fundamental to resolving the radius-albedo ambiguity in direct detection; and 4) We should consider a sustained effort on precision PRV, including investigating the limits, and then investment for a legacy duration campaign.